\documentstyle[aps,twocolumn,epsfig]{revtex}
\begin{document}
\newcommand{\ve}[1]{\mbox{\boldmath $#1$}}
\twocolumn[\hsize\textwidth\columnwidth\hsize
\csname@twocolumnfalse%
\endcsname

\draft

\title {Stability of rotating states in a weakly-interacting Bose-Einstein 
        condensate}
\author{G. M. Kavoulakis}
\date{\today}
\address{Mathematical Physics, Lund Institute of Technology, P.O. Box 118,
           S-22100 Lund, Sweden}
\maketitle

\begin{abstract}

We investigate the lowest state of a rotating, weakly-interacting
Bose-Einstein condensate trapped in a harmonic confining potential that 
is driven by an infinitesimally asymmetric perturbation. Although 
in an axially-symmetric confining potential the gas has an axially-symmetric 
single-particle density distribution, we show that in the  
presence of the small asymmetric perturbation its lowest state
is the one given by the mean-field approximation, which is a broken-symmetric
state. We also estimate the rate of relaxation of angular momentum
when the gas is no longer driven by the asymmetric perturbation
and identify two regimes of ``slow" and ``fast" relaxation. States
of certain symmetry are found to be more robust. 

\end{abstract}

\vskip0.5pc]

\section{Introduction}

  A Bose-Einstein condensate of trapped alkali-metal atoms \cite{RMP} is an 
interesting system for studying phenomena connected with superfluidity.
The behavior of this system under rotation has been studied extensively.
Experimentally vortex states in a two-component Bose-Einstein condensate
were observed by Matthews {\it et al.} \cite{JILA}. Also, Madison {\it et al.} 
\cite{Madison} and more recently Abo-Shaeer {\it et al.} \cite{KET} and
Haljan {\it et al.} \cite{Haljan} have created vortices in a single-component 
condensate.

  Theoretical studies of this problem have been performed in 
the limit of weak interactions between the atoms 
\cite{Wilkin,Rokhsar,Ben,BP,WG,LF,KMP,JKMR,Hui,LNF,Ho},
as well as in the Thomas-Fermi limit of strong interactions
\cite{Rokhsarv,FS,Emil1,GRPG,Feder,Emil2,Emil3}. Up to now most experiments
are in the Thomas-Fermi regime; however in the recent experiment
of G\"orlitz {\it et al.} \cite{Kettnew} in a cigar-shaped trap,
the limit of weak interactions was reached transversely to the 
long axis of the trap. In addition, in the experiment of
Ref.\,\cite{KET}, more than 100 vortices were created and 
as argued by Ho \cite{Ho}, in such a system one can easily get
to the regime of weak interactions. 
In the present article we restrict ourselves to the limit of 
weak interactions.

As discussed by many authors 
\cite{Wilkin,Rokhsar,Ben,BP,WG,LF,KMP,JKMR,Hui,LNF,Ho} 
in this limit there is a degeneracy due to the harmonic confining 
potential, which corresponds to the many different ways of distributing 
$L$ units of angular momentum among $N$ atoms.  The goal, therefore, 
is to incorporate the interactions which lift the degeneracy and 
determine the lowest-energy state of the system.

A basic question related to these systems concerns the lowest state 
for a given $L$ and $N$ in a symmetric confining potential.  References
\cite{Wilkin,Rokhsar,Ben,BP,WG,LF,KMP,JKMR,Hui} have dealt with 
this issue.  Having understood the question of a symmetric confining 
potential, one can then examine the ground state in the presence of a 
realistic trap, which has a small but finite asymmetry, or even in a more
deformed trap. Our main goal is to investigate (i) the single-particle
density distribution of the condensate when it equilibrates in
the rotating frame, and (ii) the characteristic timescale for 
relaxation of angular momentum of the condensate when the
trap stops to rotate and the gas is no longer driven externally.

Our article is organized in the following way: We describe
our model in Sec.\,II, examining how a small symmetry-breaking harmonic 
potential affects the lowest-energy state of a repulsive 
weakly-interacting Bose-Einstein condensate which is confined in a symmetric  
harmonic potential. In Sec.\,III we demonstrate that the state that minimizes
the energy in the presence of a small asymmetric perturbation is the
one given by the mean-field approximation, which is a broken-symmetric 
state. Finally, in Sec.\,IV we estimate the rate of relaxation
of angular momentum when the asymmetric drive stops to rotate, 
and in Sec.\,V we summarize our results.

\section{Model}
\subsection{Symmetric confining potential}

  Let us start with the Hamiltonian $H$, given by
\begin{eqnarray}
    H = \sum_{i} \left[ - \frac {\hbar^{2}} {2M} {\ve \nabla}_{i}^{2} +
  \frac 1 2 \, M \omega_{\perp}^{2} (x_{i}^{2} + y_{i}^{2}) \right],
\label{h0}
\end{eqnarray}
with $M$ being the atomic mass. In what follows we assume that
the cloud rotates around the $z$ axis, along which it remains
in its ground state, so effectively we examine a two-dimensional
problem. Let us assume for the moment that the trapping potential is that   
of an isotropic harmonic oscillator of frequency $\omega_{\perp}$ in the 
$x$-$y$ plane. For this potential the single-particle energies
$\epsilon_{n_r,m}$ are given by
\begin{eqnarray}
  \epsilon_{n_r,m} = (2 n_r + |m| + 1) \hbar \omega_{\perp},
\label{energy}
\end{eqnarray}
where $n_r$ is the radial quantum number, and $m$ is the quantum number
corresponding to the angular momentum. Let us also denote the corresponding
single-particle states as $\phi_{n_r,m}$ (single-particle states are denoted
with small initial letters and many-particle states are denoted with capital 
initial letters.) As we discuss in the following subsection,
a crucial observation according to Eq.\,(\ref{energy}) is that 
there is a degeneracy between states with ($n_r=0$; $m=m_0$), ($n_r=1$; 
$m=m_0-2$), ($n_r=2$; $m=m_0-4$), etc., with an energy $(m_0+1) \hbar 
\omega_{\perp}$. States with $m$ having the opposite sign as $L$ are excluded, 
since mixing them results in higher-energy \cite{Ben,KMP,JKMR}.

\subsection{Symmetry-breaking confining potential}

Let us now assume that in addition to the circularly-symmetric
harmonic potential there is a small symmetry-breaking potential
$\Delta H$ which has the form
\begin{eqnarray}
  \Delta H = \sum_i \frac \varepsilon 2 M \omega_{\perp}^2 (x_i^2 - y_i^2),
\label{brpot}
\end{eqnarray}
where $\varepsilon \ll 1$. The eigenvalues of the Hamiltonian $H+\Delta H$ 
are known analytically in cartesian coordinates, and they are given by
\begin{eqnarray}
    \epsilon_{n_x,n_y} = (n_x \sqrt{1 + \varepsilon} 
 + n_y \sqrt{1 - \varepsilon}) \, \hbar \omega_{\perp},
\label{energyxy}
\end{eqnarray}
where $n_x$ and $n_y$ are the number of quanta of oscillation along the 
$x$ and $y$ directions respectively. The corresponding eigenfunctions 
are given by the product of the eigenstates along the $x$ and $y$ directions
with $n_x$ and $n_y$ quanta, respectively,
\begin{eqnarray}
    \phi_{n_x,n_y}(x,y) = \phi_{n_x}(x/a_{x}) \, \phi_{n_y}(y/a_{y}).
\label{eigenxy}
\end{eqnarray}
Here $a_{i}=(\hbar/M \omega_i)^{1/2}$, is the oscillator length
along direction $i$, with $\omega_{x},\omega_{y}=\omega_{\perp} 
\sqrt{1 \pm \varepsilon}$.

Although the non-interacting problem can be solved exactly, it is 
instructive to continue working in cylindrical polar coordinates, and consider 
$\Delta H$ as a small perturbation.  In these coordinates $\Delta H$ 
links single-particle states which differ by two units of angular 
momentum, $m'-m=\pm2$.  Therefore $\Delta H$ links the degenerate 
states with ($n_r=0$; $m=m_0$), ($n_r=1$; $m=m_0-2$), ($n_r=2$; 
$m=m_0-4$), etc.  Within the subspace of these degenerate states the 
matrix elements of $\Delta H$ are all zero, except the ones linking 
neighboring states, and they are of order $\varepsilon$.  Therefore 
the problem of diagonalizing $H+\Delta H$ is equivalent to a 
tight-binding model, with a position-dependent hopping integral 
connecting only nearest neighbors.  The correction $\Delta \epsilon$ 
to the single-particle energies due to $\Delta H$ is thus of order 
$\varepsilon$, and the energy of the single-particle states is 
\begin{eqnarray}
   \epsilon_{n_r,m}^{'}= \epsilon_{n_r,m} + \Delta \epsilon
  = [m_0 + 1 + {\cal O}(\varepsilon_{n_r,m})] \, \hbar \omega_{\perp},
\label{energyc}
\end{eqnarray}
where ${\cal O}(\varepsilon_{n_r,m})$ denotes a term of order $\varepsilon$
which depends on $n_r$ and $m$. The corresponding single-particle states 
are given by
\begin{eqnarray}
  \phi_{n_r,m}^{'} = \phi_{n_r,m} + 
 \sum_{n_r^{'},m'}  {\cal O}(\varepsilon_{n_r^{'},m'})\, \phi_{n_r^{'},m'},
\label{wfc}
\end{eqnarray}
where $n_r ^{'}$ and $m'$ can take all the possible values, with the
exception $n_r^{'} \neq n_r$, and $m' \neq m$. The results of 
Eqs.\,(\ref{energyc}) and (\ref{wfc}) can also be seen by expanding 
Eqs.\,(\ref{energyxy}) and (\ref{eigenxy}) respectively in powers of
$\varepsilon$.

To be more explicit, the wavefunctions we have to consider for the 
lowest-energy state of the system are dominated by the component 
having no radial nodes,
\begin{eqnarray}
   \phi_{0,m}^{'} = \phi_{0,m} + 
 \sum_{n_r^{'},m'} {\cal O}(\varepsilon_{0,m'}) \, \phi_{n_r^{'},m'},
\label{wfcc}
\end{eqnarray}
with $n_r ^{'} =1, 2, \dots$, and $m' = m-2, m-4, \dots$
 
Equation (\ref{energyc}) implies that the lowest energy of $N$ 
(non-interacting) bosons having $L$ units of angular momentum 
(measured relative to that of the ground state) is
\begin{eqnarray}
    E(L,N) = [L + N {\cal O}(\varepsilon_{L,N})] \, \hbar \omega_{\perp},
\label{energytni}
\end{eqnarray}
where ${\cal O}(\varepsilon_{L,N})$ is a term of order $\varepsilon$ that 
depends on $L$ and $N$. According to Eq.\,(\ref{energytni}) 
there is a spreading in the energy of the states around the value
$L \hbar \omega_{\perp}$, which is of order $N \varepsilon \hbar 
\omega_{\perp}$. As long as $\varepsilon$ is small, $\varepsilon \ll 1$,
the effect of this term is not substantial for $L$ being
${\cal O} (N)$. 

\subsection{Effect of interactions -- Hierarchy of terms}

Turning to the effect of the interactions between the atoms, we 
assume that these are of zero range,
\begin{eqnarray}
  V_{\rm int} = \frac 1 2 U_{0} \sum_{i \neq j} 
 \delta({\bf r}_{i} - {\bf r}_{j}),
\label{v}
\end{eqnarray}
where $U_0 = 4 \pi \hbar^2 a_{\rm sc}/M$ is the strength of the effective 
two-body potential, with $a_{\rm sc}$ being the scattering length for 
atom-atom collisions.  In the present article we examine only the case of  
effective repulsive interactions between the atoms, $a_{\rm sc} >0$.

In the limit that the interactions are weak, the energy of the 
states which are degenerate in the absence of interactions is spread 
over a width of order $n U_0$, where $n$ is the atom density and thus 
the following condition has to be satisfied
\begin{eqnarray}
    n U_0 \ll \hbar \omega_{\perp}.
\label{cond}
\end{eqnarray}
If $N$ is the number of atoms in the trap, and assuming that the gas is 
trapped along the $z$ axis on a length scale of order $R_z$, and in the
transverse direction on a length scale of order $a_{\perp}$, $n$ is of order
$N/a_{\perp}^2 R_z$; the above condition 
can then be written as
\begin{eqnarray}
  v_0 \ll \hbar \omega_{\perp} / N,
\label{cond2}
\end{eqnarray}
where $v_0 \sim U_0/a_{\perp}^2 R_z$. Equation (\ref{cond2}) implies that 
the scattering length has to scale as
\begin{eqnarray}
  a_{\rm sc} \sim \frac {R_z} N,
\label{cond29}
\end{eqnarray}
for the assumption of weak interactions to be valid (for fixed 
$\omega_{\perp}$.) As a result, the ratio $\varepsilon \hbar \omega_{\perp}/
N v_0$ is $\ll 1$, and therefore
\begin{eqnarray}
  \varepsilon \hbar \omega_{\perp} \ll N v_0 \ll \hbar \omega_{\perp}.
\label{cond3}
\end{eqnarray}
Under the above conditions the many-body states built up by 
$\phi_{0,m}^{'}$ are quasi-degenerate. Therefore one has to deal
with a problem similar to that of 
Refs.\,\cite{Wilkin,Rokhsar,Ben,BP,WG,LF,KMP,JKMR} of a 
symmetric confining potential, with a set of (quasi)degenerate states 
whose degeneracy is lifted by the interactions.  Under the conditions 
given by Eq.\,(\ref{cond3}), the energy of the lowest state (the 
so-called yrast state) with $L$ units of angular momentum and $N$ 
particles is given by
\begin{eqnarray}
  {\cal E}(L,N) = E(L,N) + E_{\rm int}(L,N).
\label{enfin0}
\end{eqnarray}
Here $E(L,N)$ is given by Eq.\,(\ref{energytni}), and we write it for 
convenience as $E(L,N) = (L + a N \varepsilon) \hbar \omega_{\perp}$, where $a$ 
is a parameter of order unity that depends on $L$ and $N$. Also
$E_{\rm int}(L,N)$ is the interaction energy due to $V_{\rm int}$.  
For $\varepsilon = 0$, $E_{\rm int}(L,N)$ has been expressed as a power 
series in $N$ and $L$ in Ref.\,\cite{JKMR},
\begin{eqnarray}
    E_{\rm int}(L,N)  
 = (b_1 N^2 + b_2 N L + b_3 N + b_4 L + \dots) v_0,
\label{enfin}
\end{eqnarray}
where $b_i$ are parameters of order unity that depend on $L$ and $N$.
For $\varepsilon \neq 0$, $E_{\rm int}(L,N)$ becomes, to leading order 
in $\varepsilon$,
\begin{equation}
    E_{\rm int}(L,N)
 = (1 + c \varepsilon) (b_1 N^2 + b_2 N L + b_3 N + b_4 L + \dots) v_0,
\label{enfin1}
\end{equation}
where $c$ is a a parameter of order unity that depends on $L$ and $N$.
The terms proportional to $c \varepsilon$ result from the mixing of states 
with $n_r \neq 0$ (to leading order), but they are not of importance
in our problem. 

According to the above equations ${\cal E}(L,N)$ is given by (to leading order)
\begin{equation}
  {\cal E}(L,N)=
 L \hbar \omega_{\perp} + (b_1 N+ b_2 L) N v_0 + a N \varepsilon \hbar 
\omega_{\perp} + \dots,
\label{enfinn}
\end{equation}
with $L\hbar \omega_{\perp} \gg N^2 v_0 \gg N \varepsilon \hbar 
\omega_{\perp}$, so the terms of order $N^2 v_0$ in the interaction energy 
dominate the term in the energy due to the symmetry-breaking potential. For a 
given $L$, the corresponding angular frequency of rotation $\Omega$ of the gas 
is
\begin{eqnarray}
  \Omega = \frac 1 {\hbar} \frac {\partial {\cal E}(L,N)} {\partial L}.
\label{angfreq}
\end{eqnarray}
The single-particle density contours that we derive in Sec.\,III refer to a 
frame that rotates with this frequency $\Omega$, which we assume is 
also the frequency of rotation $\Omega_{0}$ of the asymmetric potential 
$\Delta H$ (under equilibrium conditions.)  In Sec.\,IV we examine the 
response of the gas if the trap stops to rotate, i.e., if 
$\Omega_{0} \rightarrow 0$.

\subsection{Connection with experiment}

   Before we proceed, we explore the relevance of our study to 
experiment. As we mentioned in the Introduction, in the experiment of 
Ref.\,\cite{KET}, more than 100 vortices were created and as 
argued in Ref.\,\cite{Ho}, in such a system one can easily 
get to the regime of weak interactions.

It is even more interesting that in the recent experiment described in 
Ref.\,\cite{Kettnew}, it has become possible to create Bose-Einstein 
condensates of $^{23}$Na atoms in cigar-shaped traps, where transversely to
the long axis of the trap the gas is in the lowest harmonic-oscillator level. 
More specifically, the trapping frequencies used were $\omega_z/2 \pi=3.5$ 
Hz and $\omega_{\perp}/2\pi=360$ Hz, which implies that $a_z \approx 11$ $\mu$m
and $a_{\perp}\approx 1$ $\mu$m. To put as many atoms as possible, G\"orlitz
{\it et al.} created a Bose-Einstein condensate which was in the Thomas-Fermi
regime of strong interactions along the long axis, but on the other hand
it was in the lowest harmonic-oscillator level perpendicular to the 
long axis. Under these conditions, the transverse width of the cloud 
is on the order of the oscillator length $a_{\perp}$, and for the 
interaction energy to be comparable to $\hbar \omega_{\perp}$,
\begin{eqnarray}
   N_c \sim \frac {R_z} {a_{\rm sc}},
\label{mna}
\end{eqnarray}
where $R_z$ is the width of the cloud along the $z$ axis.
As we mentioned, the cloud is in the Thomas-Fermi limit along this
direction, and therefore
\begin{eqnarray}
   \frac {R_z} {a_z} \approx 
  \left( \frac {N a_{\rm sc} a_z} {a_{\perp}^2} \right)^{1/3}.
\label{rz}
\end{eqnarray}
From Eqs.\,(\ref{mna}) and (\ref{rz}) we find that
\begin{eqnarray}
   N_c \sim \frac {a_z^2} {a_{\perp} a_{\rm sc}},
\label{mnas}
\end{eqnarray}
which gives $N_c \approx 5 \times 10^4$. Indeed, it was confirmed that
as $N$ varied around this value, the cloud passed from the limit of
strong interactions to the one of weak interactions.

Under the above conditions, the coherence length $\xi$ at the center of the
cloud is
\begin{eqnarray}
  \frac {\xi} {a_{\perp}} \sim \left( \frac {a_z^2} {N a_{\rm sc} a_{\perp}} 
  \right)^{1/3} \sim \left( \frac {N_c} N \right)^{1/3}.
\label{cohl}
\end{eqnarray}
For $N \ll N_c$, $a_{\perp} \ll \xi$, i.e., the coherence length
becomes larger than the size of the trap and the properties of
the system under rotation resemble those of superfluid nuclei.

\section{Density contours}

We investigate now the (normalized) single-particle density of the atoms 
$\rho({\bf r})$, in the many-body state $|\Psi_{L,N} \rangle$ with $N$ 
atoms and $L$ units of angular momentum,
\begin{eqnarray}
   \rho({\bf r}) = \frac 1 N \langle \Psi_{L,N} | \sum_i
  \delta({\bf r}-{\bf r}_i) | \Psi_{L,N} \rangle.
\label{dens}
\end{eqnarray}
From the exact solutions that come from diagonalization of $V_{\rm int}$ 
in the subspace of degenerate states \cite{Wilkin,BP}, the single-particle
density is always axially symmetric, reflecting the symmetry of the confining 
potential.

Let us now introduce a mean-field wavefunction \cite{Rokhsar,KMP}, which
is a Fock state expressed as a product of single-particle states:
\begin{eqnarray}
    \Psi_{L,N}({\bf r}_1,{\bf r}_2,\ldots,{\bf r}_N) =
  \psi({\bf r}_1) \times \psi({\bf r}_2) \ldots \psi({\bf r}_N).
\label{nwf}
\end{eqnarray}
The single-particle states $\psi({\bf r}_i)$ can be expanded
on the basis of the harmonic-oscillator eigenstates $\phi_{0,m}({\bf r}_i)$
with no radial nodes and an angular momentum $m \hbar$ along the axis of 
rotation:
\begin{eqnarray}
   \psi({\bf r}_i) = \sum_{m=0}^{\infty} c_{m} \phi_{0,m}({\bf r}_i).
\label{exp}
\end{eqnarray}
In the approach of Refs.\,\cite{Rokhsar,KMP} the coefficients $c_{m}$ in
Eq.\,(\ref{exp}) are treated as variational parameters with two
constraints imposed on them: the first is the normalization condition,
$\sum_{m}|c_{m}^{2}|=1$, and the second is that the expectation value
of the angular momentum be equal to $l=L/N$, $\sum_{m} m |c_{m}|^{2} =
L/N$.  One then calulates the interaction energy which is
\begin{eqnarray}
   E_{\rm int}(L,N) =  \frac 1 2 N^2 v_0 \int |\psi({\bf r})|^4 \,d^3r
\label{neq}
\end{eqnarray}
to leading order in $N$, and minimizes it with respect to
the $c_{m}$.  As a result of this approach, for repulsive interactions
and within the mean-field approximation a vortex array develops as
$L/N$ increases \cite{Rokhsar,KMP}.

The mean-field wavefunction above has the following
features or advantages: (i) it gives the exact leading-order term of the
interaction energy, i.e., the $N^2 v_0$ \cite{JKMR}, (ii) the corresponding
single-particle density breaks the rotational symmetry, since
the single-particle states given by Eq.\,(\ref{exp}) have some
axes of symmetry \cite{Rokhsar,KMP} by construction, and (iii)
it is essentially a one-body wavefunction, and thus it is easy to 
work with and to visualize.  As shown in Ref.\,\cite{KMP} there is a 
freedom in one of the phases of the coefficients $c_m$ reflecting the 
rotational invariance of the confining potential.  Therefore even in 
the mean-field approximation one can construct a circularly-symmetric 
state by superimposing the above states corresponding to different 
angles, and therefore to different orientations.

However, even an infinitesimally small $\varepsilon$ is enough to
change $\rho({\bf r})$ from being symmetric, looking like the one
given by mean-field, and therefore with a specific orientation.
To see this more clearly, we notice that the leading-order $N^2 v_0$ term
in the interaction energy given by the mean-field approximation is 
identical to the one that comes from the exact solution, as  
shown in Ref.\,\cite{JKMR}. For any 
nonzero value of $\varepsilon$ the term of order $N \varepsilon 
\hbar \omega_{\perp}$ has to become as small as possible, as 
Eq.\,(\ref{enfinn}) implies, in addition to the leading-order term of 
the interaction energy, i.e., the $N^2 v_0$.  In order to minimize the 
$N \varepsilon \hbar \omega_{\perp}$ term, the phase that is free to take any 
value when $\varepsilon=0$ (reflecting the rotational invariance of 
the confining potential) has to be chosen appropriately.  Therefore 
for $\varepsilon \neq 0$ there will be in general some preferred 
orientations of the cloud corresponding to some (distinct) values of 
the phase which minimize the $N \varepsilon \hbar \omega_{\perp}$ term.  These 
axes of symmetry rotate with an angular frequency $\Omega$ given by 
Eq.\,(\ref{angfreq}).

Therefore, according to our analysis, an infinitesimally asymmetric confining
potential makes the circularly symmetric single-particle density 
of the cloud unstable against the formation of a vortex array as given
by the mean-field approximation. This array of vortices has been
investigated in Refs.\,\cite{Rokhsar,KMP} (see Fig.\,2 in Ref.\,\cite{Rokhsar}
and Figs.\,3 and 4 in Ref.\,\cite{KMP}).

\section{Timescales}

Related to the above remarks are some timescales which we
examine now. We consider a problem corresponding to the experiment
of Ref.\,\cite{Madison}, where the atoms get their angular momentum
when they are still in the normal phase, above the transition
temperature, by rotating $\Delta H$ with some angular frequency
$\Omega(t)$, which slowly approaches the value $\Omega_0$.  The gas is
then cooled down to the condensed phase, in the presence of the
external rotating drive.  The system equilibrates in the rotating
frame, and eventually $\Delta H$ is turned off slowly.  After some
time the atoms are released from the trap and expand, and
their density is observed.  Since the whole process occurs
adiabatically, the state of the system is determined by minimizing the
energy in the rotating frame, i.e., by minimizing ${\cal E'}(L,N) =
{\cal E}(L,N) - L \hbar \Omega_{0}$.

Therefore initially, when the gas is in the normal phase, it has some
angular momentum, which changes during the cooling process to the
condensed phase. When the cooling process is completed the gas is not 
in equilibrium yet, since its angular frequency of rotation $\Omega$
given by Eq.\,(\ref{angfreq}) is not necessarily equal
to the frequency of rotation of the trap $\Omega_{0}$.
One therefore has to wait for some time for the system to equilibrate,
so that $\Omega$ will become equal to $\Omega_{0}$.  After this time interval
the gas is driven by the rotating asymmetric trap and it is in a
``dynamical'' steady state.

The interesting question is how does the gas respond if the
symmetry-breaking potential stops to rotate. If this
process occurs adiabatically, i.e., on timescales much larger
that $\omega_{\perp}^{-1}$, the gas will go back to rest. The
opposite limit, when the trap stops to rotate on a timescale
that is short as compared to $\omega_{\perp}^{-1}$, is non-trivial, however.
In this case the gas experiences a time-dependent potential.
This potential induces transitions to states with a different angular 
momentum. Because of the symmetry of the perturbation $\Delta H$
we have chosen (i.e., $x^2-y^2$), only states which differ by
two units of angular momentum are linked via $\Delta H$.
The rate $\Gamma$ for the system to lose two units of angular momentum (due 
to the torque coming from $\Delta H$ that no longer drives the gas) is of order
\begin{eqnarray}
    \Gamma = \frac {2 \pi} \hbar \frac
   {|\langle \Phi_{L-2,N} | \Delta H |\Psi_{L,N} \rangle|^2}
 {|\Delta {\cal E}(L,N)|}.
\label{rate}
\end{eqnarray}
Here $|\Phi_{L-2,N} \rangle$ is the state that couples to the ground 
state $|\Psi_{L,N}\rangle$ via $\Delta H$, and $\Delta {\cal E}(L,N)$
is the energy difference between them. The rate $\Gamma$
is dominated by the transition to the state that is closest in energy
to the yrast. Since the yrast state consists mostly of single-particle
states with no radial excitations, the state that is closest in energy 
is the one dominated by single-particle states with one
unit of radial excitations. The energy denominator 
in Eq.\,(\ref{rate}) is, to leading order,
\begin{eqnarray}
  |\Delta {\cal E}(L,N)| \sim N v_0,
\label{de}
\end{eqnarray}
plus corrections of order $N \varepsilon \hbar \omega_{\perp}$.

To proceed to the calculation of the matrix element of Eq.\,(\ref{rate}),
let us denote as $N_m$ the occupancy of the state of the harmonic 
oscillator $\phi_{0,m}$ when the system is in the yrast state
$|\Psi_{L,N} \rangle$. Within the mean-field approach $N_m = N |c_m|^2$,
whereas within the exact approach one needs to project the yrast state
on the basis $\phi_{0,m}$ and get the corresponding $N_m$. There is
very strong evidence that the two approaches give the same result for $N_m$
\cite{JKMR,KMR}. The above matrix element is on the order of
\begin{eqnarray}
  \langle \Phi_{L-2,N} | \Delta H |\Psi_{L,N} \rangle
 \sim \varepsilon \hbar \omega_{\perp} \sum_{m=0}^{\infty} \sqrt{N_m N_{m+2}} .
\label{me}
\end{eqnarray}
Typically $m$ in the above sum runs over a few states only, since
states with high values of $m$ are not occupied.
Now two possibilities need to be considered. Depending on the value of 
the ratio $L/N$, the sum on the right of Eq.\,(\ref{me}) can be either 
of order $N^{1/2}$, or of order $N$ \cite{Rokhsar,KMP,JKMR} and $\Gamma$
depends crucially on that. We examine each case separately below.

\subsection{Fast relaxation}

When the yrast state consists of at least two states $N_m$ and $N_{m+2}$ 
which are of order $N$, the matrix element of  Eq.\,(\ref{me}) is of
order $N\varepsilon \hbar \omega_{\perp}$ and from Eqs.\,(\ref{rate})
and (\ref{de}) we get that the rate for the system to lose $L \sim N$ 
units of angular momentum is
\begin{eqnarray}
   \frac {\Gamma} N = \omega_{\perp} \varepsilon^2 
 \frac {\hbar \omega_{\perp}} {v_0} \, \chi(L/N),
\label{deee}
\end{eqnarray}
where $\chi(L/N)$ is a function of $L/N$, and its value is of order
unity \cite{n}. 

An important conclusion resulting from Eq.\,(\ref{deee}) is that 
the timescale for relaxation of angular momentum can get long
as compared to the period of the trap, provided that 
\begin{eqnarray}
  \varepsilon \ll \left( \frac {v_0} {\hbar \omega_{\perp}} \right)^{1/2}.
\label{deeee}
\end{eqnarray}
Since, according to Eq.\,(\ref{cond2}) $v_0 \ll \hbar \omega_{\perp} / N$
for the assumption of weak interaction to be valid, therefore
\begin{eqnarray}
  \varepsilon \propto \frac 1 {N^{1/2}}
\label{dee}
\end{eqnarray}
for the decay rate to be $\ll \omega_{\perp}$. Under the
conditions of Ref.\,\cite{Kettnew}, $\varepsilon \ll 0.01$ for 
$N= 5 \times 10^4$, which is a rather restrictive condition.
Realistic symmetric traps have anisotropies on the order of 1\% \cite{Dalib},
and therefore the relaxation rate of angular momentum should be on the
order of the trap frequency, even for the least asymmetric traps.

The above case of fast relaxation applies to all the yrast states with 
$L/N \le 2.03$, except for the region around the single vortex, $L/N \approx 1$ 
\cite{KMP}. Above the value $L/N \approx 2.03$ the system shows a discontiuous 
transition from a two-fold symmetric state to a three-fold symmetric state
\cite{KMP}. Also for  $L/N \approx 1$ the yrast state is dominated by the
occupancy of a single state, the $m=1$ \cite{Wilkin,Rokhsar,Ben,KMP}.
These cases are examined in the following subsection.

\subsection{Slow relaxation}

For $L/N \agt 2.03$, the yrast state is three-fold symmetric;
therefore for $L/N \agt 2.03$ and for $L/N \approx 1$,  
the matrix element of Eq.\,(\ref{me}) is instead given by
\begin{eqnarray}
  \langle \Phi_{L-2,N} | \Delta H |\Psi_{L,N} \rangle
 \sim \sqrt{N} \varepsilon \hbar \omega_{\perp}.
\label{mes}
\end{eqnarray}
In this case,
\begin{eqnarray}
   \frac {\Gamma} N \sim \omega_{\perp} \varepsilon^2
 \frac {\hbar \omega_{\perp}} {N v_0}.
\label{deees}
\end{eqnarray}
From Eq.\,(\ref{deees}) one sees that for the state to be stable,
\begin{eqnarray}
  \varepsilon \ll \left( \frac {N v_0} {\hbar \omega_{\perp}} \right)^{1/2},
\label{deeees}
\end{eqnarray}
which is a condition much easier to satisfy, since although $\varepsilon$
has to be numerically small, it does not have to scale with the number
of atoms, $N$. 

To summarize the results of this section, the presence of
$\Delta H$ that rotates with an angular frequency $\Omega_0$ results in
(i) transferring angular momentum, and (ii) favouring the creation of a
vortex array instead of a circularly-symmetric ground state.  Finally
stopping the rotation of $\Delta H$ results in a relaxation rate of the 
angular momentum of the gas that depends on the yrast state considered, 
i.e., on the ratio $L/N$. For $L/N \le 2.03$, except $L/N \approx 1$, the 
rate is low as long as $\varepsilon \sim N^{-1/2}$ and the states are stable 
under such weak perturbations. For $L/N \approx 1$, and for $L/N \agt 2.03$,  
the relaxation rate is suppressed by a factor of $1/N$, and $\varepsilon$ 
has to be numerically small, but independent of $N$, for these states to 
be stable. 

The relaxation rate can also be evaluated for higher values
of $L/N$. According to Ref.\,\cite{Rokhsar}, these states have
certain symmetry, and as long as this symmetry is not two-fold \cite{fold},
the relaxation rate is given by Eq.\,(\ref{deees}).

\section{Conclusions}

In conclusion, we have investigated the effect of a small 
asymmetry in the confining potential of a weakly-interacting 
Bose-Einstein condensate, which rotates bringing the gas into rotation. 
Three important results have come out of this study: (i) the limit of
weak interactions can be achieved, even under current experimental conditions,
(ii) even in a weakly-asymmetric rotating trap, in its lowest state the system
develops a vortex array for an effective repulsive interaction between the 
atoms, and (iii) if the gas is no longer driven externally, the rate of
relaxation of the angular momentum depends on the yrast state considered. For
the specific form of the perturbation $\Delta H$ (i.e., $x^2-y^2$), if
the yrast state does not have some symmetry, or if it has a two-fold symmetry,
the rate of relaxation of angular momentum is large. On the other hand,
the unit vortex, or states with $s$-fold symmetry, with $s \ge 3$ are expected 
to be more robust against anisotropies of the confining potential, since the  
rate of relaxation of angular momentum is lower by a factor of $1/N$.
These are definite theoretical predictions and it 
is interesting to confirm them experimentally after the expected creation 
of vortices in weakly-interacting Bose-Einstein condensates.

\vskip2pc

\centerline{\bf ACKNOWLEDGMENTS}

\vskip1.0pc

I am grateful to A. Jackson, B. Mottelson, and S. M. Reimann
for useful discussions.


\begin{references}
\bibitem{n} For a noninteracting gas, $v_0=0$, the denominator in the 
   above equation $|\Delta {\cal E}(L,N)|$ is $\sim N \varepsilon \hbar 
   \omega_{\perp}$, (which was previously neglected because of the 
   condition $N \varepsilon \hbar \omega_{\perp} \ll N v_0$), and 
   $\Gamma/N \sim |\varepsilon| \omega_{\perp}$.
\bibitem{fold} We note that for an $s$-fold symmetric state, only
   the states $N_m$ are macroscopically occupied, where $m=\alpha s$ 
   with $\alpha = 0,1,2,\dots.$
\end{references}
\end{document}